\documentclass[useAMS,usenatbib]{mn2e}
\usepackage{amssymb,amsmath,epsfig,times,natbib,multirow,booktabs}

\newcommand{\swift}{{\it Swift}}

\newcommand{\plm}{$\pm$}

\long\def\symbolfootnote[#1]#2{\begingroup%
\def\thefootnote{\fnsymbol{footnote}}\footnote[#1]{#2}\endgroup} 

\title[The X-ray Low State in Mrk~1048]{A Partial Eclipse of the Heart: The Absorbed X-ray Low State in Mrk~1048}

\author[M. L. Parker et al.]{M. L. Parker,$^1$\thanks{Email: mlparker@ast.cam.ac.uk}
  N. Schartel,$^2$
  S. Komossa,$^3$
  D. Grupe,$^{4,5}$
  M. Santos-Lle\'o,$^2$\newauthor
  A. C. Fabian$^1$
  and S. Mathur$^{6,7}$\\
  $^{1}$Institute of Astronomy, Madingley Road, Cambridge, CB3 0HA\\
  $^{2}$XMM-Newton Science Operations Centre, ESA, Villafranca del Castillo, Apartado 78, 28691 Villanueva de la Canada, Spain\\
  $^{3}$Max-Planck-Institut f{\"u}r Radioastronomie, Auf dem H{\"u}gel 69, D-53121 Bonn, Germany\\
  $^{4}$Swift Mission Operation Center, 2582 Gateway Dr., State College, PA 16801, USA\\
  $^{5}$Space Science Center, Morehead State University, 235 Martindale Dr., Morehead, KY 40351, USA\\
  $^{6}$Department of Astronomy, Ohio State University, 140 West 18th Avenue, Columbus, OH 43210, USA\\
  $^{7}$Cosmology and Astro-Particle Physics, Ohio State University, Columbus, OH 43210, USA
}
\date{}
\begin{document}

\maketitle

\begin{abstract}
We present two new \emph{XMM-Newton} observations of an unprecedented low flux state in the Seyfert~1 Mrk~1048 (NGC~985), taken in 2013. The X-ray flux below 1~keV drops by a factor of 4--5, whereas the spectrum above 5~keV is essentially unchanged. This points towards an absorption origin for the low state, and we confirm this with spectral fitting, finding that the spectral differences can be well modelled by the addition of a partial covering neutral absorber, with a column density of $\sim3\times 10^{22}$~cm$^{-2}$ and a covering fraction of $\sim0.6$.
The optical and UV fluxes are not affected, and indeed are marginally brighter in the more recent observations, suggesting that only the inner regions of the disk are affected by the absorption event. This indicates either that the absorption is due to a cloud passing over the inner disk, obscuring the X-ray source but leaving the outer disk untouched, or that the absorber is dust-free so the UV continuum is unaffected. We use arguments based on the duration of the event and the physical properties of the absorber to constrain its size and location, and conclude that it is most likely a small cloud at $\sim10^{18}$~cm from the source.
\end{abstract}


\section{Introduction}
Low flux states in active galactic nuclei (AGN) are of particular interest as they can show very extreme behaviour, such as drops by factors of 10 or more in flux over the 2--10~keV band  on time scales of months to days \citep[e.g.][]{Grupe07}. The nature of these low states may have important implications for our understanding of the behaviour of AGN in general, and may be relevant to measurements of spin.

One of the key questions for any such low state is whether the drop in flux is intrinsic to the source \citep[as in 1H~0707-495;][]{Fabian12} or can be attributed to intervening material \citep[as in NGC~1365;][]{Risaliti13}. In the former case, observations of low states in AGN may allow use to probe the very inner edge of the accretion disk and make very accurate measurements of the spin \citep[Parker et al., submitted]{Fabian14}, while in the latter these observations can be used to observe the environment surrounding AGN, including outflows \citep{Kaastra14}, broad line region (BLR) clouds and the edge of the torus \citep{Maiolino10,Lohfink12}, and constrain the size of the X-ray emitting corona \citep{Sanfrutos13}. Unfortunately, these two different causes for low-flux states can be hard to distinguish \citep[e.g.][]{Schartel07}, due to low count spectra and intrinsically similar spectral shapes. Interestingly, a dramatic but more gradual decline in the luminosity (including the X-ray luminosity) from Mrk~509 appears to have coincided with a change in the Seyfert classification of the source \citep{Denney14}. 

Mrk~1048 (also known as NGC~985) is a Seyfert~1 galaxy \citep{VeronCetty83} at redshift 0.0427. The galaxy itself has a clear ring structure, which suggests that it is going through a merger \citep{deVauc75}. The presence of warm absorption in the X-ray spectrum of this source is well known, and was first suggested based on \emph{ROSAT} observations by \citet{Brandt94}. These features were then confirmed using \emph{ASCA} by \citet{Nicastro99}, and have also been seen with grating spectra by \emph{Chandra} and \emph{XMM-Newton} \citep{Krongold05,Krongold09}. In this work we present \emph{XMM-Newton} observations of an unusual low flux state in Mrk~1048, detected using \emph{Swift} monitoring.

\section{Observations and Data Reduction}

In an ongoing search for AGN in deep low-states \citep[e.g.][]{Schartel10}, we noticed the low Swift flux of Mrk 1048, and then triggered our dedicated \emph{XMM-Newton} follow-ups. We use an archival \emph{XMM-Newton} observation from 2003, when the source was not in a low state, for comparison with the new data.

\subsection{XMM-Newton}

Mrk~1048 was observed three times with \emph{XMM-Newton} \citep{Jansen01}, once in 2003 and twice in 2013. The details of the observations are given in Table~\ref{obstable}. 
We used the \emph{XMM-Newton} Science Analysis System (SAS) v13.5.0 for all data reduction, using the \textsc{epproc} task to produce cleaned \emph{EPIC}-pn \citep{Struder01} event files. \textsc{evselect} was then used to extract spectra, filtering for background flares and using only single and double events. Response matrices were generated using the \textsc{rmfgen} and \textsc{arfgen} tasks. Source and background spectra were extracted from 40" circular regions for all observations, and all spectra are source-dominated until $>10$~keV. The background regions are selected to avoid contaminating sources and copper lines from the detector. The \emph{RGS} data were reduced using the \textsc{rgsproc} ftool, and filtered for background flares.
All spectral fitting is performed using Xspec version 12.8.1l \citep{Arnaud96}, and the spectra are binned to a minimum of 30 counts per bin using \textsc{grppha}. We use $\chi^2$ statistics for the fits presented in this work, but we also checked the validity of our fits using Cash statistics \citep[\textsc{cstat} in \textsc{xspec};][]{Cash79} and find no significant differences in the acceptability of the fits or in the values of model parameters. The Optical Monitor (OM) data was extracted using \textsc{omichain}, and corrected for Galactic reddening using the reddening curve of \citet{Cardelli89}.

\begin{table*}
\centering
\begin{tabular}{c c c c c c}
\hline
Obs. ID & Start Date & Exposure time & Count Rate & 0.3--2~keV Flux & 2--10~keV Flux\\
& & (ks) & (s$^{-1}$) & erg~cm$^{-2}$~s$^{-1}$ & erg~cm$^{-2}$~s$^{-1}$\\
\hline
0150470601 & 2003-07-15 & 31.14 & $3.94\pm0.01$ & $5.68\times10^{-12}$ & $9.17\times10^{-12}$  \\
0690870101 & 2013-07-20	& 15.92	& $1.64\pm0.01$	& $1.57\times10^{-12}$ & $8.12\times10^{-12}$  \\
0690870501 & 2013-08-10	& 71.62	& $1.763\pm0.05$& $1.91\times10^{-12}$ & $7.96\times10^{-12}$  \\
\hline
\end{tabular}
\caption{Details of the three \emph{XMM-Newton} observations used in this analysis. The exposure time and count rate are for the \emph{EPIC}-pn detector after filtering for background flares and background subtraction. Fluxes are corrected for Galactic absorption.}
\label{obstable}
\end{table*}

\begin{figure*}
\includegraphics[width=14cm]{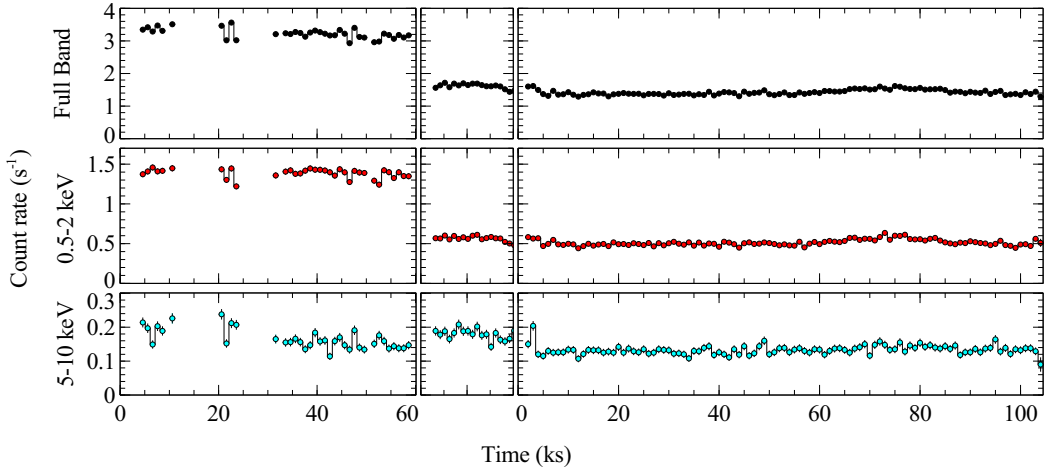}
\caption{\emph{EPIC}-pn lightcurves for the three \emph{XMM-Newton} observations of Mrk~1048. The first column shows the data from the 2003 `normal' flux state observation, and the other two columns show the short and long 2013 low state observations. The three rows show the full band, 0.5--2 and 5--10~keV lighcurves, respectively. Data are binned into 1~ks intervals. Gaps in the first observation are due to background flares.}
\label{lightcurves}
\end{figure*}

\begin{figure}
\includegraphics[width=\linewidth]{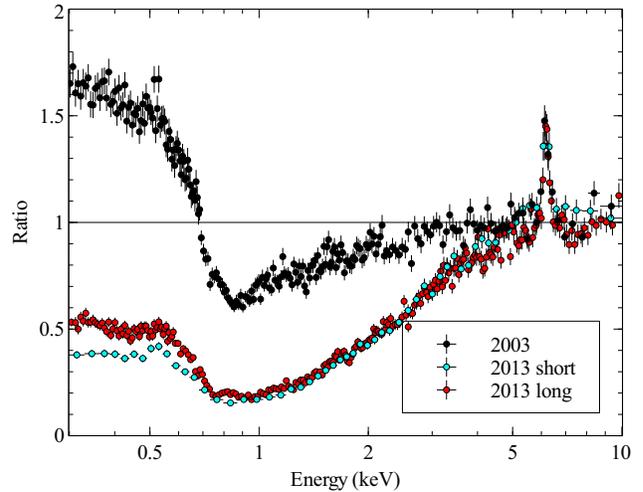}
\caption{Ratios of the \emph{EPIC}-pn spectra of the three \emph{XMM-Newton} observations of Mrk~1048 to the same (in both normalisation and index) absorbed powerlaw, fit from 5--10~keV, where the spectra are very similar. Data are rebinned in Xspec for clarity.}
\label{powerlawratios}
\end{figure}

In Fig.~\ref{lightcurves} we show the \emph{EPIC}-pn lightcurves for Mrk~1048. There is a large drop (a factor of $\sim3$) in flux between the 2003 observation and the 2013 observations, which is concentrated in the 0.5--2~keV band, with almost no changes in flux over the 5--10~keV band. We show the three spectra in Fig.~\ref{powerlawratios} as ratios to a powerlaw, modified by galactic absorption, fit from 5--10~keV. It is clear from this figure that almost all of the variability takes place at low energies, and a narrow iron line and warm absorption are clearly visible.

\subsection{Swift}
Mkn 1048 was a target of a \swift\ fillin program between 2007 July and 2008 June \citep{Grupe10} and was observed 5 times during that period. When Swift observed Mkn 1048 again on 2013-July-07 it discovered Mkn 1048 in a low X-ray flux state. consequentially we requested further \swift\ observations between 2013-July 12 and August 11. These observations are listed in Table\,\ref{swift_log}.

All observations with the \swift\ X-ray Telescope \citep[XRT,][]{Burrows05} were performed in photon counting mode (Hill et al., 2005). XRT data were reduced with the the task {\it xrtpipeline} version 0.12.6., which is included in the HEASOFT package 6.12. Source counts were collected in a circular region with a radius of 94.3$^{''}$ and the background in a close-by source-free region with a radius of 282.9$^{''}$.
The auxiliary response files (ARFs) were created by the FTOOL {\it xrtmkarf} and we used the most recent response file {\it swxpc0to12s6\_20010101v013.rmf}. All spectra were rebinned with 20 counts per bin, except for the data in segments 008 and 012 which were binned with one count per bin, due to the low number of total counts in these observations. These spectra were fitted in {\it XSPEC} using Cash statistics, and all others were fitted with $\chi^2$ statistics. 

The photometric data taken by the UV-Optical Telescope \citep[UVOT,][]{Roming05} were analyzed with task {\it uvotsource}. Source counts were selected in a circular region with a radius of 5$^{''}$ and a nearby source-free region with a radius of 20$^{''}$. Count rates were converted into fluxes and magnitudes based on the most recent UVOT calibration as described in \citet{Poole08} and \citet{Breeveld10}. The UVOT data were corrected for Galactic reddening of $E_{\rm B-V}=0.033$ \citep{Schlegel98}. The correction factor in each filter was calculated with equation (2) in \citet{Roming09} who used the standard reddening correction curves by \citet{Cardelli89}.

\begin{table}
\centering
\begin{tabular}{ccccc}
\hline
		&			&				& \multicolumn{2}{c}{Exposure time (s)}\\
\cmidrule(r){4-5}
ID 		& Segment 	& Start Date 	& XRT 	& UVM2\\
\hline
91616	& 001		& 2013-07-07	& 2148	& 526\\
36530	& 006		& 2013-07-12	& 2400	& 582\\
		& 007		& 2013-07-20	& 1051	& 247\\
		& 008		& 2013-07-25	& 1049	& 1038\\
		& 009		& 2013-07-29	& 952	& 962\\
		& 010		& 2013-08-02	& 967	& 954\\
		& 012		& 2013-08-10	& 974	& 981\\
		& 013		& 2013-08-11	& 849	& 199\\
\hline

\end{tabular}
\caption{\swift\ observation log of Mrk~1048. From the \emph{UVOT} data we state only the UVM2 exposure times, as the other filters were not used for some observations.}
\label{swift_log}
\end{table}

\begin{figure}
\includegraphics[width=\linewidth]{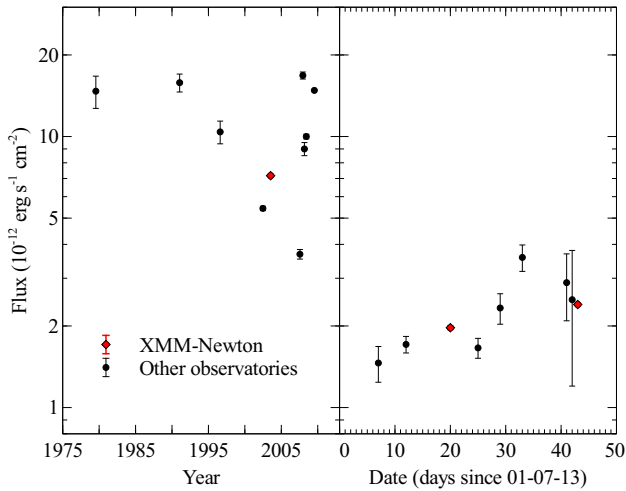}
\caption{Long term 0.2--2~keV light curve of Mrk~1048, showing the 2013 low state. The three \emph{XMM-Newton} observations are plotted as red diamonds. The left panel shows all the data up to 2013, and the right panel shows the measurements from \emph{XMM-Newton} and \emph{Swift} in July and August 2013.}
\label{longtermlcurve}
\end{figure}

In Fig.~\ref{longtermlcurve} we show the long term lightcurve of Mrk~1048, which has been observed by every major X-ray mission. The source flux has never been observed to drop as low as in the 2013 low state, which is around 20 times fainter than Mrk~1048 at its brightest. The additional data used in this plot comes from \emph{EINSTEIN} \citep{Fabbiano92}; \emph{ASCA} \citep{Nicastro99}; \emph{Suzaku} \citep{Winter12}; \emph{Swift} \citep{Grupe10} and \emph{Chandra} \citep{Krongold05}.

\section{Spectral Analysis}

\subsection{2003 XMM-Newton data}
\label{2003section}
We initially focus on the data from the 2003 observation, with the aim of establishing a baseline model from which to examine the changes in the low flux state observations of 2013.

\begin{figure*}
\includegraphics[width=8.5cm]{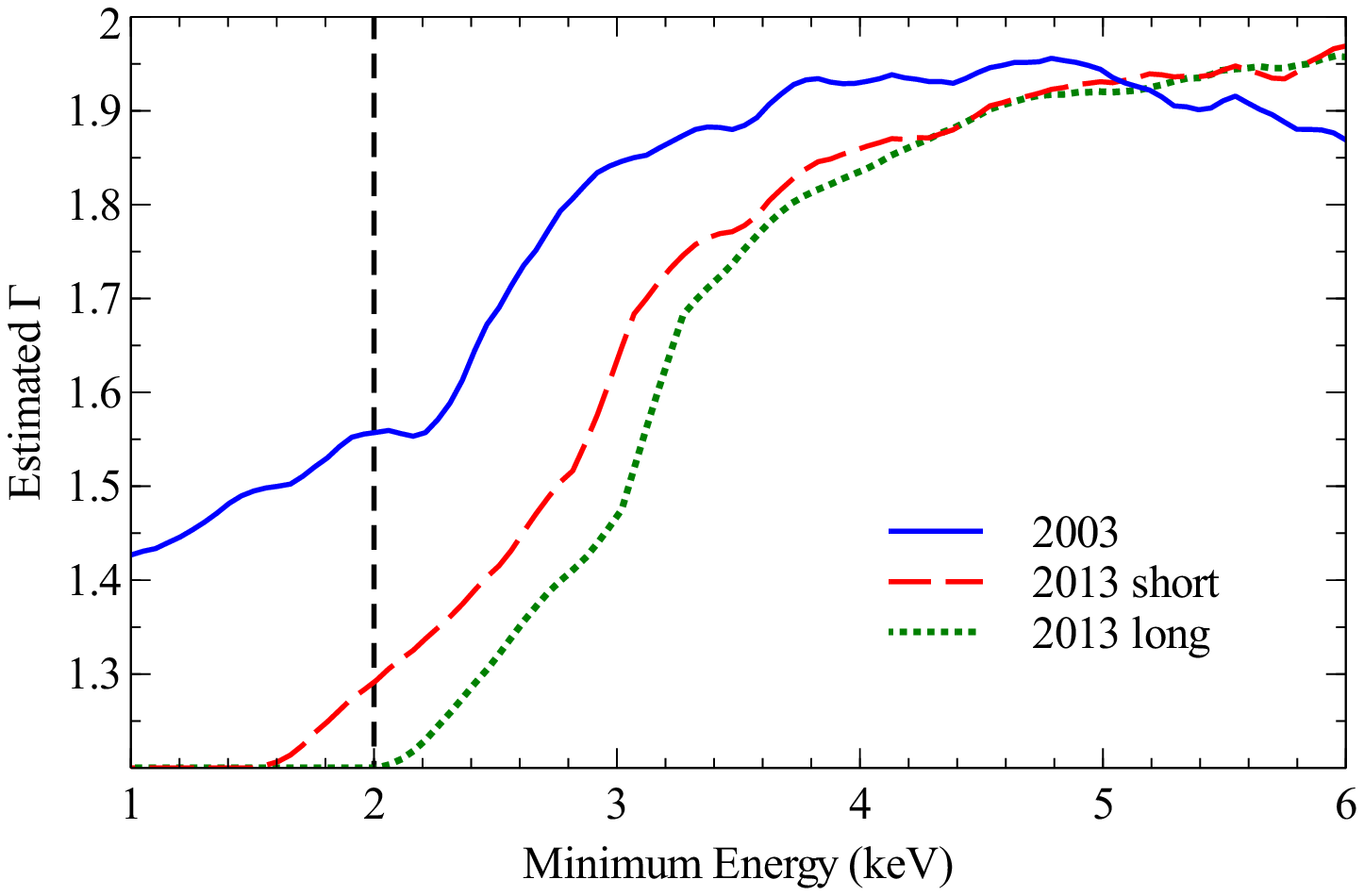}
\includegraphics[width=8.5cm]{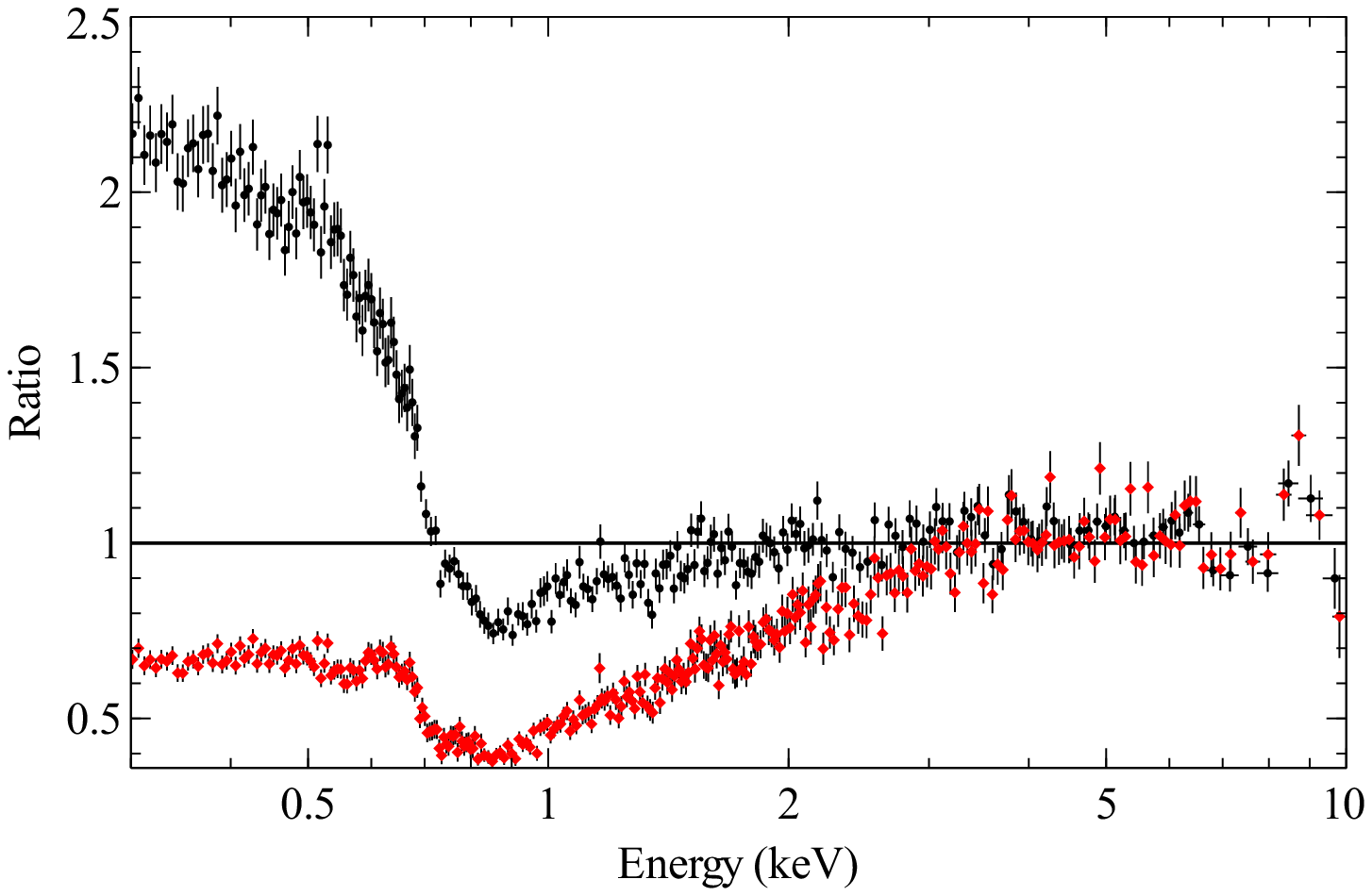}
\caption{Left: Estimated photon index $\Gamma$ as a function of the energy band used to fit the data, for the three different observations. The upper limit is set to 10~keV in all cases, while the lower energy limit is varied. The dashed line shows the conventional 2--10~keV band. Right: Data to model ratio for absorbed power law plus distant reflection fits over the 2--10~keV (black) and 5--10~keV (red) energy bands, extrapolated to lower energies for the 2003 \emph{EPIC}-pn data. The soft excess is replaced with a deficit at low energies when the energy band is restricted to higher energies.}
\label{gammaestimates}
\end{figure*}

In previous work analysing the 2003 data, the \emph{EPIC}-pn data are first fit from 2--10~keV, and the fit is then extrapolated down to lower energies, showing clear evidence of warm absorption and a soft excess. However, as we show in Fig.~\ref{gammaestimates}, the estimated power law index from such a fit is highly dependent on the energy band chosen, and this can have a very large effect on the extrapolated low energy ($<2$~keV) power law flux. For the fitting in Fig.~\ref{gammaestimates}, we use a simple power law plus distant reflection model, including galactic absorption, using \textsc{xillver} \citep{Garcia13} with $\log(\xi)=0$ to model the distant reflection. The full Xspec model used is \textsc{tbabs*(powerlaw+xillver)}, and we fix the column density of the galactic absorption to $3\times 10^{20}$~cm$^{-2}$ \citep{Kalberla05}.
As shown in the figure, a much harder powerlaw index is obtained when this model is fit from 2--10~keV than from 5--10, which indicates that there is still spectral curvature from the warm absorption present above 2~keV.
Based on the flattening of the plots above $\sim3$~keV and the consistency between the different observations, it seems likely that the powerlaw index should be steeper than that found from the 2--10~keV band, around 1.9. Extrapolating such a power law model to lower energies does not show a low energy excess, rather it leaves a deficit which could potentially be explained by the warm absorption (Fig.~\ref{gammaestimates}, right panel). This plot shows the impact of extrapolating the continuum from two different energy bands. The model extrapolated from the 2--10~keV band shows a very strong soft excess, whereas no such feature is present when the model is extrapolated from the 5--10~keV band. The indices for these fits are $1.5\pm0.1$ and $1.9\pm0.2$, respectively. Because of this result, we do not include a separate soft excess component in our fits.

To investigate the full spectrum (from 0.3--10~keV) we use an \textsc{xstar} grid to fit the warm absorption. A fit with a single warm absorption zone (\textsc{tbabs*(xstar*powerlaw+xillver+zgauss)}) gives an excellent fit ($\chi^2_\nu=1099/1094$), and this is not significantly improved by the addition of a second absorbing zone. The parameters of this fit are shown in the first row of Table~\ref{fitpars}. An additional narrow line is included (modelled with \textsc{zgauss}) at 6.55~keV, as a significant excess is found in the fit to all observations (see \S~\ref{section_allfits}), with its parameters fixed at the best fit values from the joint fit to all 3 observations.


\begin{figure}
\includegraphics[width=\linewidth]{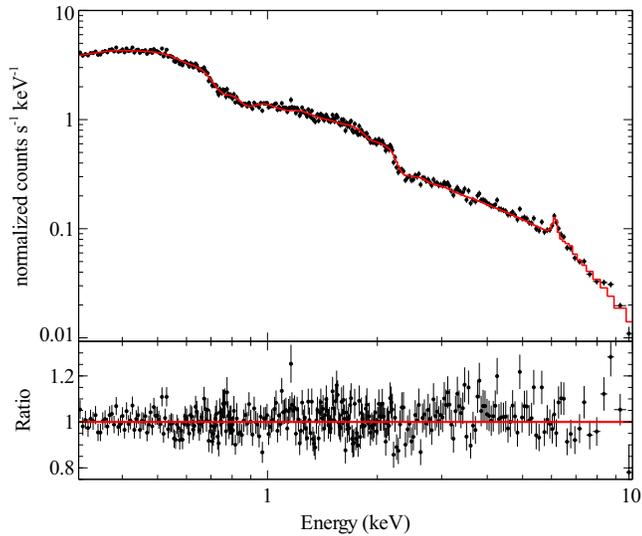}
\caption{Top: Data from the 2003 high flux state observation of Mrk~1048, fit with a power law plus distant reflection model, modified by warm absorption and galactic absorption. Bottom: Ratio of the data to the model.}
\label{2003fits}
\end{figure}

\subsection{Differences between observations}

\begin{figure}
\includegraphics[width=\linewidth]{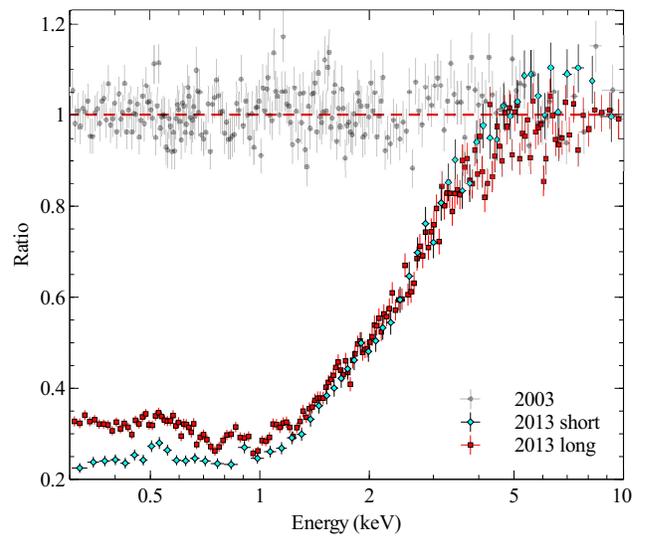}
\caption{Ratio of the \emph{EPIC}-pn spectra of the two 2013 observations to the best fit model of the 2003 data. This variability is dominated by changes below $\sim4$~keV, consistent with absorption from neutral material. There are also two potential absorption lines visible around 0.8 and 1~keV in the 2013 long observation. Data are rebinned in Xspec for clarity}
\label{ratiotoobs1}
\end{figure}

For the purposes of comparing the observations, we first plot the 2013 low state observations as a ratio to the best-fit model of the 2003 data (Fig.~\ref{ratiotoobs1}). For this plot, the 2003 model can be effectively regarded as a phenomenological fit. The figure clearly shows that the spectral changes take place almost exclusively below 5~keV, with a sharp drop down to $\sim1$~keV where it flattens again. This spectral shape, with low and high energy breaks and no strong features, is exactly that predicted by partial covering neutral absorption, and we therefore focus on modelling this interpretation.

\begin{figure}
\includegraphics[width=\linewidth]{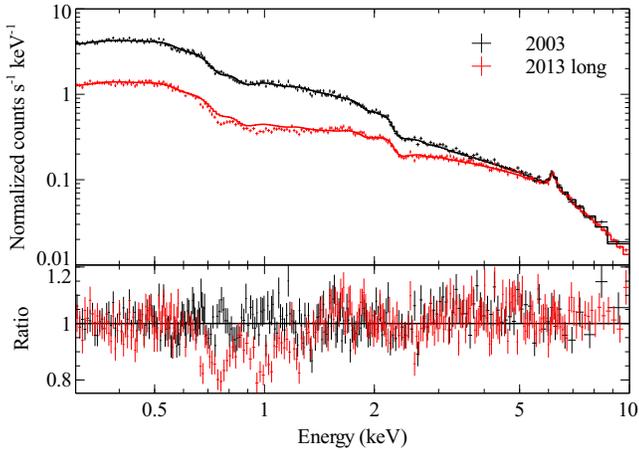}
\caption{Data and ratio for a joint fit to the 2003 and 2013 long observations, excluding the 0.5--1.5~keV band in 2013. For this fit we allow only changes in neutral absorption between the two observations. Strong residual features are visible at $\sim 0.95$ and $0.75$~keV.}
\label{wabschanges}
\end{figure}

To investigate potential changes in the warm absorber(s), we simultaneously fit the 2003 and 2013 long observations with the best fit model discussed in \S~\ref{2003section}. To this model, we add a partial covering neutral absorber, modelled with \textsc{pcfabs} in Xspec. The covering fraction and column density are fixed at zero in the 2003 spectrum, and free to vary in the 2013 observation. We exclude the 0.5--1~keV band from the 2013 data, where there are small absorption features visible in Fig.~\ref{ratiotoobs1}, then jointly fit the data. The resulting residuals are shown in the lower panel of Fig.~\ref{wabschanges}, showing clear evidence of changes in the ionised absorption.

\begin{figure}
\includegraphics[width=\linewidth]{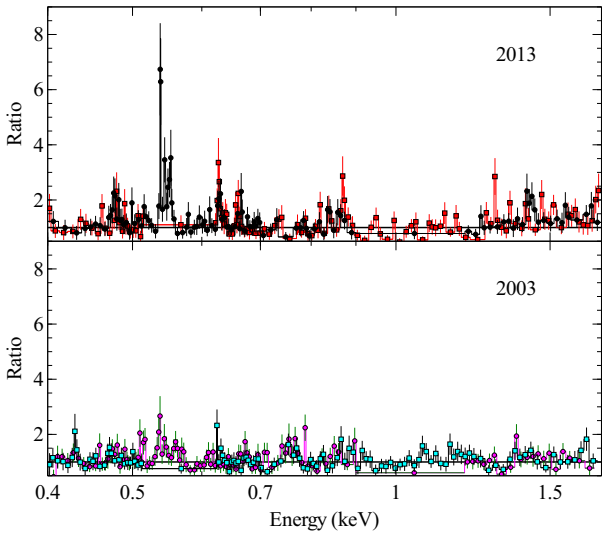}
\caption{Emission lines remaining after fitting a powerlaw continuum modified by warm absorption and cold partially covering absorption, for the 2003 and 2013 \emph{RGS} data. The strongest lines are the \textsc{Ovii} triplet from 0.56--0.57~keV, which is the strongest line feature predicted by the distant reflection model (see Fig.~\ref{bestfit}).}
\label{rgsplot}
\end{figure}

\begin{table}
\centering
\begin{tabular}{l c c}
\hline
Line  & Energy (eV)& Flux (10$^-14$~erg~s$^{-1}$~cm$^{-2}$)\\
\hline
\textsc{Ovii} triplet & 561 & $3.3^{+1.2}_{-0.1}$\\
 & 569 & $6.6^{+2.4}_{-1.9}$\\
 & 574 & $5.0^{+2.4}_{-1.6}$\\
\textsc{Oviii} K$_\alpha$ & 653 & $1.8^{+1.0}_{-0.5}$ \\
\hline
\end{tabular}
\caption{Fluxes of the narrow lines found in the \emph{RGS} data from the low flux state observation.}
\label{linefluxes}
\end{table}

While detailed modelling of the \emph{RGS} data is beyond the scope of this work, we can use the spectra to investigate potential changes in the absorption and emission between the observations. In Fig.~\ref{rgsplot} we show the data/model ratios for the 2003 observation and the 2013 long observation, where both spectra are fit with the full model including partial covering and warm absorption (see \S~\ref{section_allfits}), but with the distant reflection component removed. Strong narrow emission lines are clearly visible in the 2013 data, and are either not detected or much weaker in the 2003 spectrum. The strongest emission feature here is the \textsc{Ovii} triplet, which is also the strongest feature predicted in this band by the distant reflection model (Fig.~\ref{bestfit}, right panel). These lines do not correspond to strong absorption features in the model (with the exception of the feature above 0.7~keV in the 2003 data, which corresponds to the iron UTA), and therefore are genuine emission features. This suggests that while the primary continuum is highly absorbed, the reflection component is unaffected, leaving much more prominent emission lines. 

Fitting narrow Gaussian lines to the low flux spectrum at the energies of the four lines identified by \citet{Krongold09} in the high state returns line fluxes which are consistent with those found by \citeauthor{Krongold09}, although the errors are large (Table~\ref{linefluxes}). This supports the conclusion that the distant reflection has not changed between the observations, and certainly has changed much less than the continuum.

\subsection{Fits to all observations}
\label{section_allfits}
To investigate the spectral changes in detail, we simultaneously fit the \emph{EPIC-pn} spectra from all three observations. In all the models discussed here, we find that an additional narrow iron line is required to properly model the high energy spectrum. The typical energy for this feature (from the best fit model) is $6.55\pm0.03$~keV, and the equivalent width is 38~eV. We do not find significant variability in the strength or energy of the line either between observations or between models. 

As expected, based on the large drop in flux (Fig.~\ref{ratiotoobs1}) and only small changes in the absorption lines (Fig.~\ref{wabschanges}), a model without an additional cold absorber (allowing changes in the ionisation and column density of the warm absorber only) gives a poor fit ($\chi^2_\nu=6048/3120=1.93$). Similarly, allowing only the continuum and reflection to vary (both in spectral index and normalisation) between the observations also gives a poor fit ($\chi^2_\nu=4850/3123=1.55$). Allowing both the underlying spectrum and warm absorption to change gives a much better fit ($\chi^2_\nu=3692/3122=1.18$), although significant residuals remain around 0.8--0.9~keV, where a strong oxygen absorption edge is present in the ionised absorption model, but absent in the data. Adding additional warm absorption zones does not significantly improve this fit, as the (very small) drop in $\chi^2$ is offset by the drop in degrees of freedom.

\begin{table*}
\centering
\begin{tabular}{c c c c c c c c c c c}
\hline
&\multicolumn{2}{c}{Neutral absorber}&\multicolumn{2}{c}{Warm absorber}& Powerlaw & \multicolumn{2}{c}{Distant reflection}\\
\cmidrule(r){2-3} \cmidrule(r){4-5}\cmidrule(r){6-6} \cmidrule(r){7-8}
Model & $n_\textrm{H}$	& Covering Fraction& $n_\textrm{H}$	& $\log(\xi)$& $\Gamma$ & $A_\textrm{Fe}$ & $\theta$ &$\chi^2/$d.o.f.\\
& $(10^{22}$cm$^{-2})$&& $(10^{22}$cm$^{-2})$ & (log[erg~cm~s$^{-1}$])&  &(solar)& (degrees) \\
\hline
1 &&& $2.36\pm0.07$ &$1.951\pm0.005$ &  $1.96\pm0.01$ & $<0.53$ & $73\pm5$ & 1099/1094\\
\\
2a & & & $2.17\pm0.06$ & $1.952\pm0.006$ & $1.94\pm0.01$ & $<0.585$ & $82\pm1$ & 3629/3122\\
2b & & & $4.78\pm0.12$ & $1.942\pm0.006$ & $1.50\pm0.02$   &-&-&-&\\
2c & & & $4.72\pm0.06$ & $1.940\pm0.003$ & $1.677\pm0.007$ &-&-&-&\\

\\
3a & 0* & 0* & $2.32\pm0.05$ & $1.937\pm0.004$ & $1.961^{+0.003}_{-0.004}$ & $<0.51$ & $64\pm2$ & 3423/3127\\
3b & $2.47\pm0.07$ & $0.788\pm0.003$ & - & -&-&-&-&-&\\
3c & $3.02\pm0.06$ & $0.718\pm0.002$& - & -&-&-&-&-&\\

\\
4a & 0* & 0* & $2.35\pm0.07$ & $1.952\pm0.005$ & $1.959\pm0.006$ & $<0.51$ & $75_{-3}^{+1}$ & 3165/3118\\
4b & $2.82_{-0.2}^{+0.3}$ & $0.66\pm0.03$ & $2.6\pm0.3$ & $1.91\pm0.01$&$1.82\pm0.02$&-&-&-&\\
4c & $4.2\pm0.2$ & $0.59\pm0.02$& $2.4\pm0.1$ & $1.78\pm0.02$& $1.89\pm0.01$&-&-&-&\\
\hline
\end{tabular}
\caption{Fit parameters for the models discussed in \S~\ref{2003section} and \S~\ref{section_allfits}. Values left blank are not used in a given model, and those marked with `-' are fixed between observations. Model 1 is the best fit to the 2003 (high state) observation only. Model 2 is the fit to all three spectra, allowing for changes in the photon index and warm absorption between observations. Model 3 ties the warm absorption and underlying spectrum between observations, but allows for a variable cold absorber, and model 4 allows the warm absorber and powerlaw to change as well. a,b and c correspond to the three observations in Table~\ref{obstable}. Errors are not shown for the covering fraction and column density in 3a and 4a, as their low values and degeneracy mean that they cannot be independently constrained (see text).}
\label{fitpars}
\end{table*}

\begin{figure*}
\includegraphics[width=8.5cm]{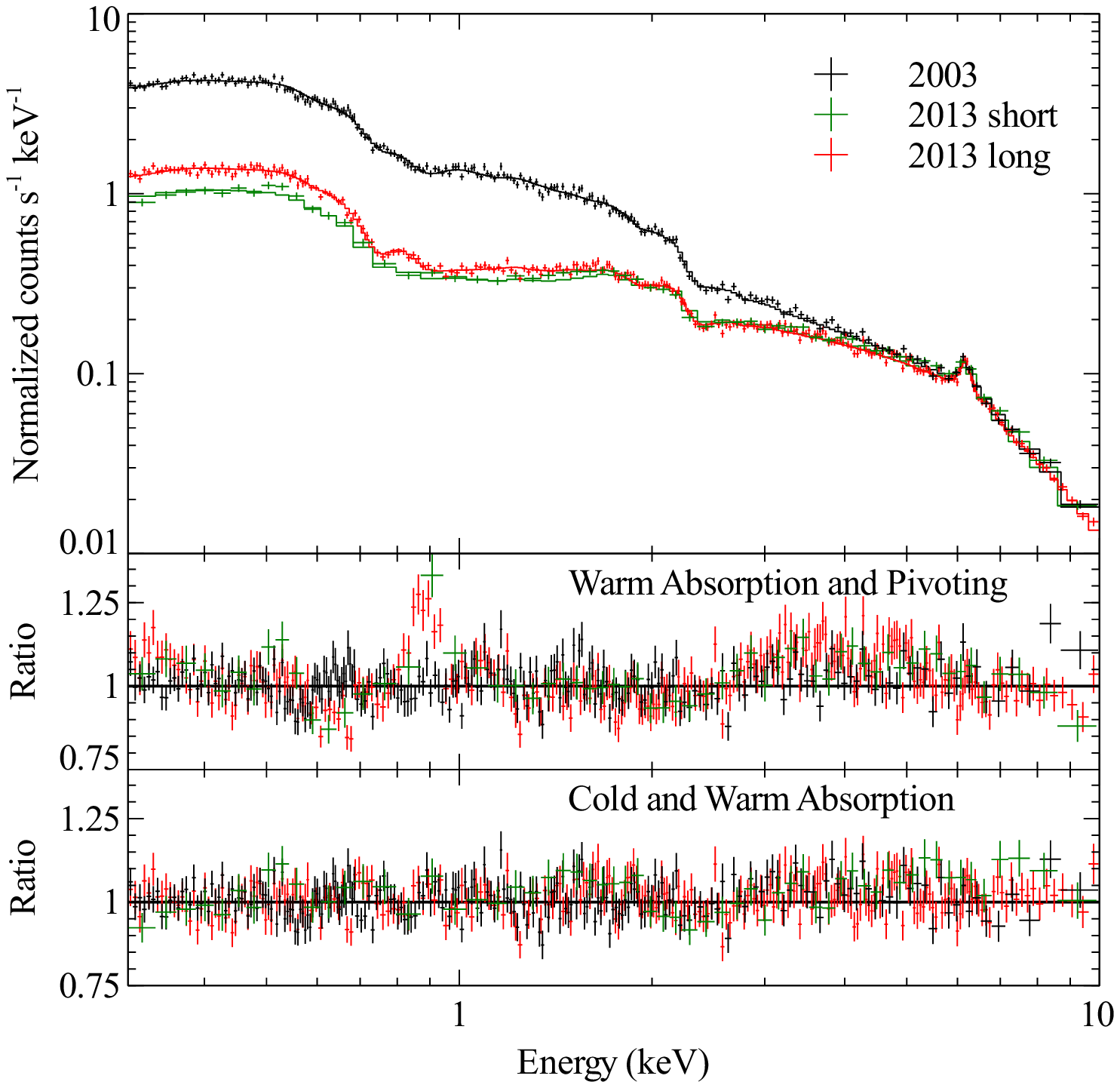}
\includegraphics[width=8.5cm]{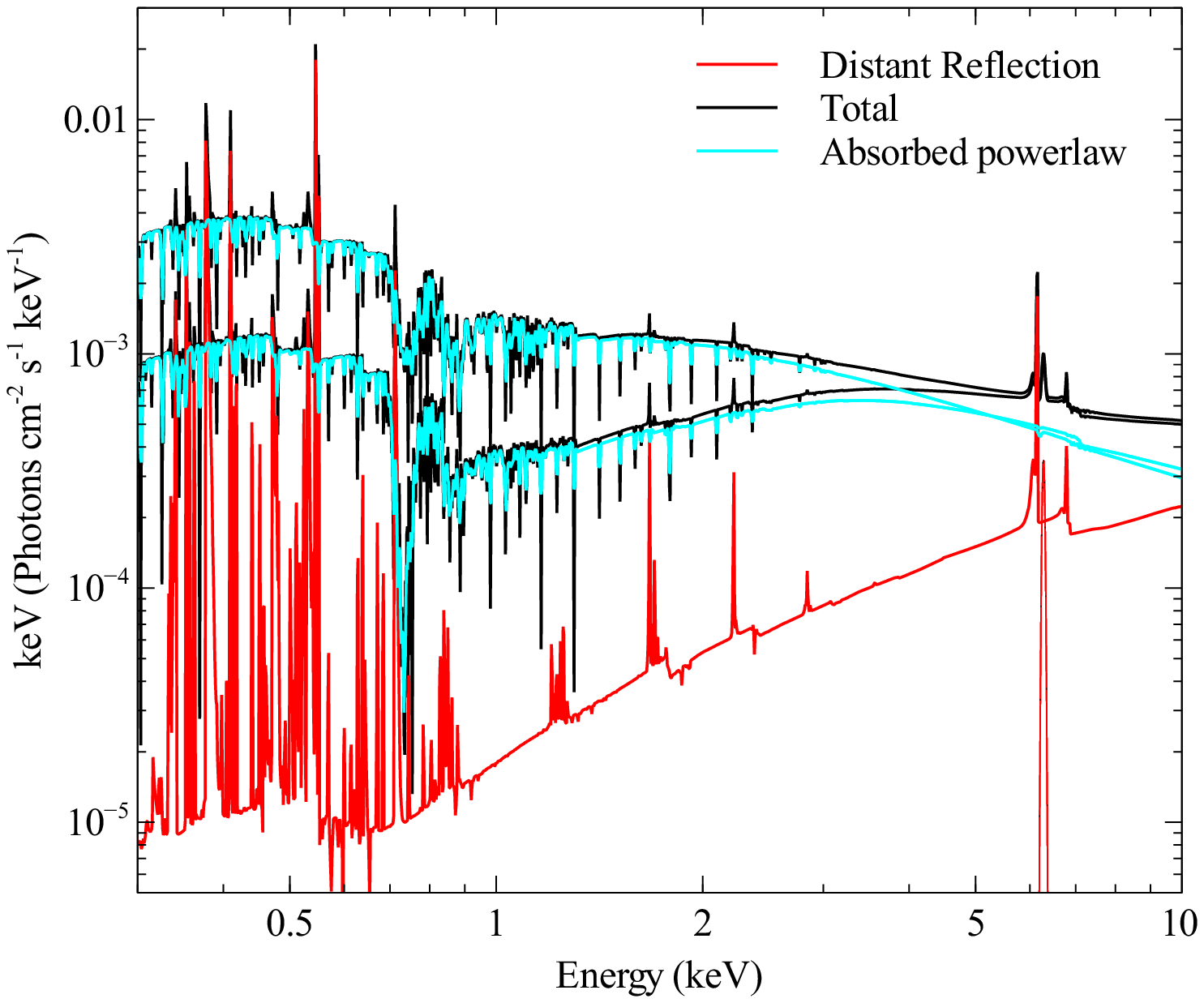}
\caption{Left: The data and residuals to the best fit model for all three observations. Data has been binned in Xspec for clarity. Right: The models fit to the 2003 and 2013 long observations, showing the absorbed powerlaw and distant reflection components, including the narrow 6.55~keV line. In all cases, the upper line corresponds to the 2003 observation. The models have been smoothed slightly using a moving average for clarity.}
\label{bestfit}
\end{figure*}

We next investigate the addition of a cold absorber, modelled using \textsc{zpcfabs} (full model: \textsc{tbabs * (zpcfabs * xstar * powerlaw + xillver + zgauss)}). Keeping all parameters apart from the covering fraction and column density tied between the observations, we find a much improved fit over the spectral pivoting models ($\chi^2_\nu=3423/3127=1.09$), with a $\delta\chi^2$ of -206 for 5 more degrees of freedom. However, as shown in Fig.~\ref{wabschanges} there do appear to be changes in the ionised absorption between the 2003 and 2013 observations, once cold absorption is taken into account. For our final best-fit model we allow the column density and ionisation of the warm absorber to change between observations, along with the photon index and flux of the power law. This gives a very good fit ($\chi^2_\nu=3165/3118=1.02$), with no significant residual features. The bulk of the spectral changes are caused by the cold absorption, with only minor changes needed in the continuum and warm absorption. We show the fits to all three datasets in the left panel of Fig.~\ref{bestfit}, both with and without the cold absorber. The best fit model is shown in the right panel of Fig.~\ref{bestfit}. Based on this model, after correcting for both Galactic and intrinsic absorption, we find X-ray luminosities of $L_\textrm{X, 0.3--2}=3.85\times10^{43}$~erg~s$^{-1}$ and $L_\textrm{X, 2--10}=4.21\times10^{43}$~erg~s$^{-1}$.

\section{Optical Monitor and Swift Data}
\label{sec_omuvot}

The X-ray and UV fluxes from \swift\ monitoring are shown in Table~\ref{swift_res}. We show only the UWM2 flux, as the other filters were not used for half of the 2013 observations. The XRT flux shows a very large (a factor of $\sim$17) drop from 2008 to 2013, but the UV flux actually increases over that interval. Over the course of the 2013 monitoring, the UV flux is very stable, with only small changes in flux, whereas the X-ray flux varies considerably (a factor of $\sim$2.5).

\begin{table}
\begin{center}
\begin{tabular}{clrc}
\hline
ID 		& Segment 	& $F_{\rm 0.2-2.0 keV}$  	& $F_\textrm{UVM2}$\\
\hline
36530	& 001$^{1}$	& 3.68$^{+0.21}_{-0.16}$	& ---\\
		& 003$^{1}$	& 17.00$^{+0.55}_{-0.80}$	& 57.1\plm1.5 \\
91616	& 001		& 0.99\plm0.10				& 39.0\plm1.2 \\
36530	& 006		& 1.43$^{+0.12}_{-0.10}$	& 40.8\plm1.2 \\
		& 007		& 0.99$^{+0.16}_{-0.20}$	& 38.1\plm1.2 \\
		& 008		& 1.26$^{+0.14}_{-0.44}$	& 38.1\plm1.2 \\
		& 009		& 1.72$^{+0.21}_{-0.44}$	& 41.1\plm1.2 \\
		& 010		& 2.51$^{+0.28}_{-0.50}$	& 43.5\plm1.2 \\
		& 012		& 1.95$^{+0.66}_{-0.55}$	& 41.4\plm1.2 \\
		& 013		& 1.70$^{+1.03}_{-0.40}$	& 40.8\plm1.2 \\
\hline

\end{tabular}
\caption{\swift\ observed XRT 0.2-2.0 keV and UVOT UVM2 fluxes. All fluxes are given in units of $10^{-12}$ ergs s$^{-1}$ cm$^{-2}$ and the UV fluxes are corrected for Galactic reddening.
$^{1}$These observations were previously published in \citep{Grupe10}, and are from 23 July and 17 December 2007. We list these here for comparison purposes. Note that the UVOT magnitudes differ from those given in Grupe et al. (2010) due to changes in the calibration.}
\label{swift_res}
\end{center}
\end{table}

This is confirmed by the OM data (Table~\ref{OMdata}), which shows a significant increase in UV flux in the 2013 observations over the 2003 observation.

\begin{table*}
\centering
\begin{tabular}{c c c cc c c}
\hline
	& \multicolumn{3}{c}{UV Flux} & \multicolumn{3}{c}{Optical Flux}\\
\cmidrule(r){2-4} \cmidrule(r){5-7}
Obs.	& UVW1	& UVW2	& UVM2	& U	& B	& V\\
\hline
0150470601	& $37.31\pm0.06$	& $38.1\pm0.8$	& $36.65\pm0.13$	& -& - &-\\
0690870101	& $43.10\pm0.03$	& $46.4\pm0.7$	& $38.86\pm0.08$	& $35.12\pm0.03$ & $25.25\pm0.05$ & $32.8\pm0.1$\\
0690870501	& $45.137\pm0.009$ & $48.5\pm0.7$ & $40.49\pm0.08$	& $37.87\pm0.03$ & $26.69\pm0.05$ & $33.9\pm0.1$\\
\hline
\end{tabular}
\caption{\emph{XMM-Newton} OM data for Mrk~1048. All fluxes are in units of $10^{-12}$~erg~s$^{-1}$~cm$^{-2}$, and are corrected for Galactic reddening.}
\label{OMdata}
\end{table*}

\section{Discussion}

While it is clear that multiple models can fit the underlying spectrum of this source (e.g. with and without a separate soft excess), this does not change our conclusions about the dominant cause of spectral variability between 2003 and 2013. The model used for the underlying spectrum can effectively be treated as a phenomenological fit, until more detailed data or modelling can distinguish the different models. The spectral differences will still be well modelled with partial covering neutral absorption, regardless of the intrinsic spectrum.
While the high-state \emph{RGS} data from 2003 were well-suited to a two-component warm absorber model \citep{Krongold09} the low state data do not provide additional constraints on the absorption features, but reveal strong emission features. These lines appear to be consistent in flux with those found by \citeauthor{Krongold09}, and appear much more prominent due to the reduced continuum flux. The constancy of these features implies that they are caused by reflection outside the region being obscured, potentially from a dusty torus or gas in the broad or narrow line regions.

Unlike other observations of AGN low states Mrk~1048 does not show any features of relativistic reflection. Several sources show strong, broad iron lines in these low states \citep[e.g. PG~2112+059 and 1H~0707-495:][]{Schartel10, Fabian12}, and a recent NuSTAR observation of Mrk~335 in a low flux state shows both a strong broad iron line and Compton hump (Parker et al., submitted). We do not see any evidence for relativistic reflection in any of the observations presented here, and unlike the other sources the drop in flux in Mrk~1048 is strictly confined to low energies. 

The OM and UVOT data presented in \S~\ref{sec_omuvot} do not show the large drop in flux seen in the X-ray spectrum of Mrk~1048 in 2013. This suggests that only the inner regions of the disk are affected by this event. Similarly, the \swift\ monitoring in 2013 reveals that the X-ray flux is much more variable than the UV throughout this period. As the majority of the flux variability in the X-ray band appears to be driven by neutral absorption, this suggests that the cold absorption identified here is not affecting the UV or optical flux. There are two possible explanations for this: firstly, the absorption could be caused by a small cloud or clouds passing in front of the disk, blocking the emission from the compact X-ray emitting region \citep{Dai10} but not impacting flux from the much larger UV and optical emitting regions; alternatively, if the absorber is dust-free there will be no extinction of the UV continuum, and the absorber could be much larger.

In the following, we use the results of the analysis shown in this paper to better constrain the location and size of the absorber. With this aim, we use the observed duration of the obscuration event (at least 40 days), the limited variations of the absorption parameters during the event (Table~\ref{fitpars} fit 4b and 4c), and the values of those parameters. In particular, we use the non-unity covering fraction (0.6--0.7) and the column density (2.5--4.4$\times10^{22}$~cm$^{-2}$). We also use the fact that the UV emission is not affected by the absorption event, and the low ionization (consistent with being neutral) of the obscuring material.

The absorption event in 2013 clearly lasted for at least 40 days, based on the long-term light curve (Fig.~\ref{longtermlcurve}), with only small variations in the covering fraction and column density. This suggests a single large absorber or multiple small clouds passing over the disk, as a single small cloud should cause more rapid variability. However, the effect of a large cloud or multiple small clouds passing over the disk might be expected to be noticeable in the UV band, unless they were dust free. 
Another issue is the covering fraction. A large absorber (i.e. one significantly larger than the X-ray emitting region) would be expected to fully cover the source.
If we assume, based on the non-unity covering fraction, that the absorbing cloud is around the same size as the X-ray emitting region, we can calculate a lower limit on the orbital radius of the cloud. We assume a black hole mass of $2\times10^8 M_\odot$ \citep{Kim08,Vasudevan09} and a diameter of 10~$R_\textrm{G}=3\times10^{14}$~cm for the corona \citep[based on microlensing constraints, e.g.][]{Dai10} and the same for the cloud diameter, $D_\textrm{c}$. Using the equation for the total orbit duration in terms of the cloud size and Keplerian velocity from \citet{Risaliti07}, we find that in order to have an eclipse that lasts longer than 1 month a small cloud must be located at a radius $r_\textrm{c}\gtrsim8\times10^{17}$~cm from the source. This corresponds to $\sim3\times10^4R_\textrm{G}$, which is consistent with a cloud in the broad line region.

For a larger cloud, where the cloud diameter is significantly greater than that of the source, the equation from \citet{Risaliti07} simplifies such that the eclipse time scales as $t\propto r_\textrm{c}^{1/2}D_\textrm{c}$. We can calculate an upper limit on where such a cloud could lie based on the dust sublimation radius\footnote{We note that this estimate assumes that the cloud originated at larger radii - if the cloud had come from inside the sublimation radius, it could have formed dust-free and moved outwards.}, as the cloud must be dust free in order to not obscure the UV emission. Assuming a bolometric correction factor $K=20$ \citep{Vasudevan07} based on the low Eddington ratio \citep[$\lambda_\textrm{Edd}=0.02$,][]{Vasudevan09} and an ionizing flux equal to half the bolometric flux \citep[$L_\textrm{bol}=10^{44.8}$~erg s$^{-1}$,][]{Vasudevan09}, we find a dust sublimation radius of $r_\textrm{sub}\sim6\times10^{17}$~cm using equation 5 from \citet{Barvainis87}. This radius is only marginally smaller than the radius inferred for a cloud of size similar to the X-ray emitting region. Therefore, for a cloud at this radius, the required size to produce an eclipse with a duration greater than 40 days is only marginally larger than the X-ray source, and partial covering of the source is plausible.

A final constraint on the location of the obscuring medium comes from the ionization of the gas \citep[e.g.][]{Risaliti07}. By replacing the neutral partial covering absorber with an ionized one \citep[modelled using \textsc{zxipcf},][]{Reeves08}, we calculate an upper limit on the ionization of $\log(\xi)<-0.62$. From the definition of $\xi$, this implies a minimum radius of $r_\textrm{c}=(L/n\xi)^{1/2}\sim3\times10^{18}$~cm for a cloud with $n$, the density of the gas, given by $n=n_\textrm{H}/D_\textrm{c}$ (see Table~\ref{fitpars}). This distance scales as $D_\textrm{c}^{1/2}$, so a larger absorber would have to be further out. However, this constraint is inconsistent with the requirement that a larger cloud lies within the dust sublimation radius, so we conclude that the cloud is most likely to be small (around 10$^{14}$--10$^{15}$~cm) and at a radius $r_\textrm{c}\gtrsim10^{18}$~cm from the source. We note that the density inferred for such a cloud, $n= n_\textrm{H}/D_\textrm{c}\sim 3\times10^{22}/3\times10^{14}=10^{8}$~cm$^{-3}$, is low compared to that generally assumed for BLR clouds (10$^9$--10$^{11}$~cm$^{-3}$) or observed in other sources \citep[e.g.][]{Risaliti11}. One possible explanation for this is that we are not viewing the source through the dense core of the cloud, rather through the more diffuse outer regions \citep[see e.g.][]{Maiolino10}. This should not greatly affect our estimate of the cloud location or size, as the dust sublimation constraint is independent of the cloud size, the ionization constraint will still apply to the outer regions of the cloud, and partial covering becomes implausible if the cloud is a great deal larger.

Mrk~1048 has a very unusual morphology, with a prominent ring structure \citep{deVauc75} and a gas-rich double nucleus \citep{PerezGarcia96}. It is interesting to speculate that this remarkable morphology and recent merger may be responsible for absorption of the AGN, by allowing cold gas to fall in towards the nucleus. 

We note that there appears to be a small emission line at around 9~keV in the spectrum of the 2003 observation (Fig.~\ref{2003fits} and~\ref{gammaestimates}). This feature also appears to be present in the spectra shown in \citet{Krongold09}, although it is difficult to see as the residuals are plotted in standard deviations, rather than as a ratio. The origin of this feature is unclear, as we have selected our source and background regions to avoid contamination from the copper emission in the detector. Regardless, this feature is too small to have a significant impact on any of the results and conclusions based on the data from this observation.

\section{Conclusions}

We have presented new \emph{XMM-Newton} observations of a low flux state in Mrk~1048. The spectra show a drop in flux by a factor of $\sim5$ below 1~keV compared to an archival observation from 2003, while above 5~keV the spectrum is unaffected. 

We find that the source spectrum can be well modelled with a single warm absorption zone, applied to a power law continuum and distant reflection. Based on fits to different energy bands we find that the large soft excess identified in previous work is a model dependent result. The soft excess can be well modelled by warm absorption and distant reflection when the continuum is carefully extrapolated from bands unaffected by absorption.

The drop in flux can be described simply by the addition of a partial covering neutral absorber, with some additional minor changes in the warm absorption. Reflection dominated models can be discounted by the lack of relativistic reflection features and the total absence of variability above 5~keV, and low states driven by intrinsic spectral changes, warm absorption, or both are ruled out by spectral fitting. 

We conclude that the low flux state in Mrk~1048 is caused by obscuration of the inner disk by cold gas and find that this is most likely due to an obscuring cloud at a radius of a few $10^{18}$~cm from the source.

\section*{Acknowledgements}
The authors would like to thank the anonymous referee for their detailed and helpful comments, and Dom Walton for the \textsc{xstar} grids used in this work.
MLP acknowledges financial support from the Science and Technology Facilities Council (STFC), and would like to thank Ranjan Vasudevan and Matt Middleton for helpful discussions.
This work is based on observations obtained with XMM-Newton, an ESA science mission with instruments and contributions directly funded by ESA Member States and NASA. 
At Penn State we acknowledge support from the NASA Swift program
through contract NAS5-00136.
\bibliographystyle{mnras}
\bibliography{bibliography_mrk1048}
\end{document}